\documentclass{article}

\usepackage{arxiv}

\usepackage[utf8]{inputenc} % allow utf-8 input
\usepackage[T1]{fontenc}    % use 8-bit T1 fonts
\usepackage{hyperref}       % hyperlinks
\usepackage{url}            % simple URL typesetting
\usepackage{booktabs}       % professional-quality tables
\usepackage{amsfonts}       % blackboard math symbols
\usepackage{nicefrac}       % compact symbols for 1/2, etc.
\usepackage{microtype}      % microtypography
\usepackage{lipsum}
\usepackage{graphicx}
\graphicspath{ {./images/} }
\usepackage{amsmath}

\title{A Unified Model for Voice and Accent Conversion
in Speech and Singing Using Self-Supervised
Learning and Feature Extraction}

\author{
 Sowmya Cheripally \\
 Department of Computer Science and Engineering \\
 BVRIT Hyderabad College of Engineering for Women \\
 Hyderabad, Telangana, 500090, India \\
  \texttt{19wh1a05g9@bvrithyderabad.edu.in} \\
}

\begin{document}
\maketitle
\begin{abstract}
This paper presents a new voice conversion model capable of transforming both speaking and singing voices. It addresses key challenges in current systems, such as conveying emotions, managing pronunciation and accent changes, and reproducing non-verbal sounds. One of the model’s standout features is its ability to perform accent conversion on hybrid voice samples that encompass both speech and singing, allowing it to change the speaker’s accent while preserving the original content and prosody. The proposed model uses an encoder-decoder architecture: the encoder is based on HuBERT to process the speech’s acoustic and linguistic content, while the HiFi-GAN decoder audio matches the target speaker’s voice. The model incorporates fundamental frequency (f0) features and singer embeddings to enhance performance while ensuring the pitch \& tone accuracy and vocal identity are preserved during transformation. This approach improves how naturally and flexibly voice style can be transformed, showing strong potential for applications in voice dubbing, content creation, and technologies like Text-to-Speech (TTS) and Interactive Voice Response (IVR) systems.
\end{abstract}

% keywords can be removed
\keywords{Accent Conversion \and Hifi-GAN \and HuBERT \and Singing Voice Conversion \and Speech Synthesis \and Voice Conversion}

\section{Introduction}
Recent advancements in speech synthesis have produced high-quality, understandable speech. Even said, a lot of approaches still fall short of fully capturing the speaker's unique style and emotional attributes. This issue is particularly apparent in tasks that require these qualities, including voice conversion. 

\begin{figure}  % 't!' forces the figure to the top of the column
    \centering
    \includegraphics[width=0.8\linewidth]{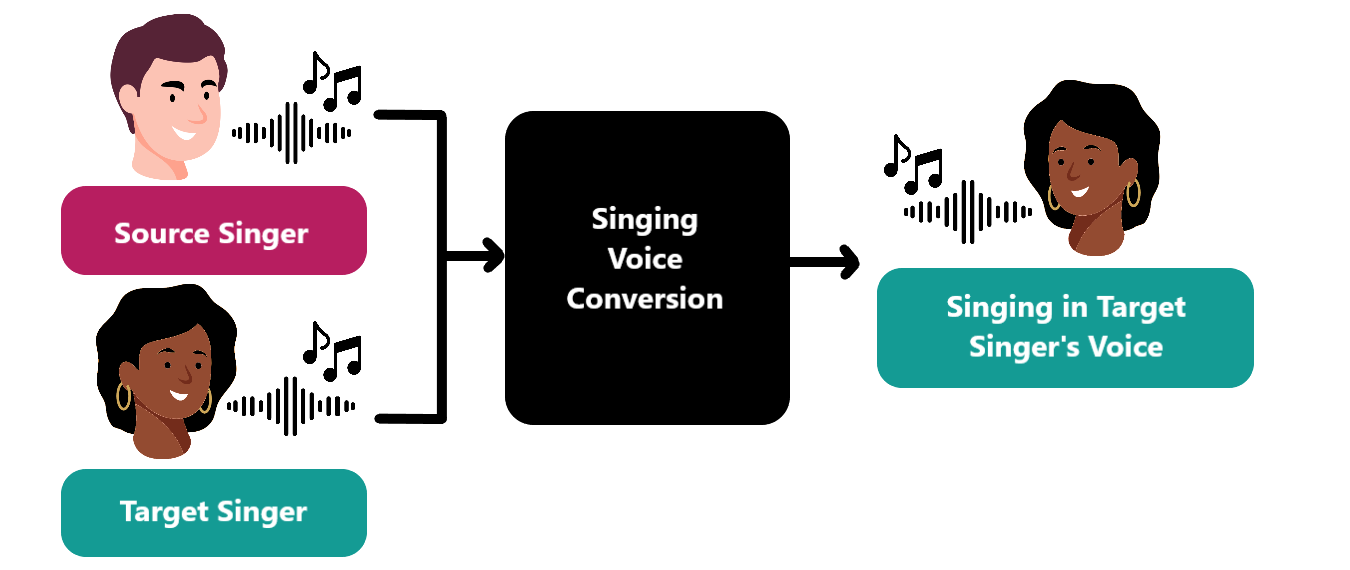}  % Adjust image width as needed
    \caption{
        Representation of Singing Voice Conversion
    }
    \label{fig1}
\end{figure}

\begin{figure}  % 't!' forces the figure to the top of the column
    \centering
    \includegraphics[width=0.8\linewidth]{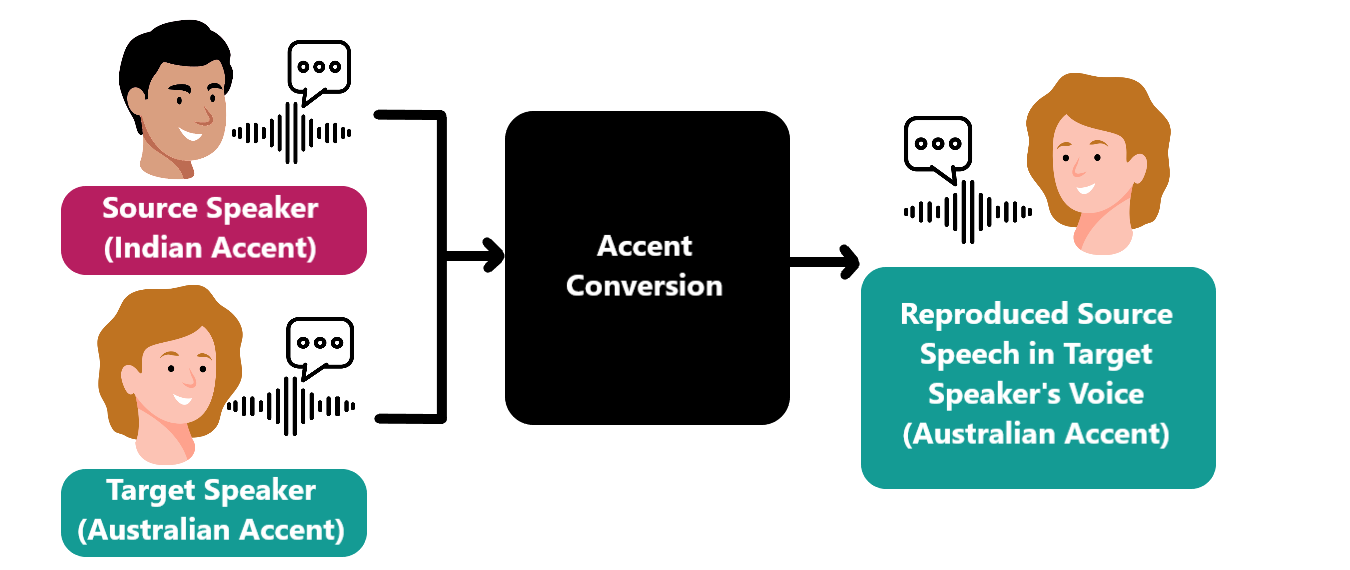}  % Adjust image width as needed
    \caption{
        Representation of Accent Conversion
    }
    \label{fig2}
\end{figure}

Voice style transfer is a technique for making the speech of one speaker sound like that of a different speaker. Singing voice conversion (SVC) extends the definition of normal VC (another term for voice style transfer) and aims to convert the singing voice of a source singer to match that of a target singer without changing its contents. Similar to SVC, accent conversion transforms the accent of a speech to make it sound like another accent while preserving the linguistic content. 

In previous work, a voice conversion system combining the WaveNet Vocoder and VAE-GAN was presented, demonstrating its success in generating speech and transferring lexical content [1]. However, the model struggled with synthesizing non-verbal sounds, conveying emotions, and transferring accents, which remains a significant limitation in current technology. 

This paper introduces an improved voice conversion model to overcome these limitations.  Our proposed system integrates several components: an f0 feature module, a singer embedding module, a self-supervised feature extractor, and a HiFi-GAN decoder. This innovative model extracts fundamental frequency and feature embeddings from the source audio, processes them using a HuBERT-based speech encoder, and generates high-quality audio in the target singer’s voice. Including f0 scaling and shifting during inference further enhances the model’s ability to match the source and target voices accurately. 

Since this approach has shown to be capable of producing natural and adaptable results, this model has far-reaching applications. These include voiceover and dubbing for actors, multilingual voice adaptation, privacy protection, and content generation across various media. The model’s ability to refine speech synthesis technologies like TTS and Interactive Voice Response (IVR) accentuates its potential impact on industry and society. 

\section{Literature Survey}
\label{sec:literaturesurvey}
Early endeavors in voice conversion, such as those by Kayte et al. [2] and Percybrooks et al. [3], utilized Hidden Markov Models (HMMs) to convert voices without requiring explicit phonetic labelling, achieving improvements in speech quality and speaker identity conversion.  

In later developments, non-parallel voice conversion emerged as a more flexible alternative to parallel methods, which depended heavily on paired data of the same utterances from different speakers—datasets that were typically scarce and difficult to obtain. 

Concurrently, the renewed interest in neural network architectures—particularly neural vocoders like WaveNet [4], WaveRNN [5], and HiFi-GAN [6]—has sparked further exploration of voice conversion techniques, enabling more efficient and high-fidelity speech synthesis. 

Deng et al. proposed PitchNet [7], which introduced a pitch adversarial network to enhance the precision of pitch manipulation in unsupervised singing voice conversion models. This improvement enabled more accurate pitch translation and flexible control over the converted singing voice, addressing issues with out-of-key singing that had plagued earlier models. 

More recently, AlBadawy and Lyu presented a model that combines Variational Autoencoders (VAE) and GANs for speech-to-speech neuro-style transfer. This model demonstrated strong performance across various metrics but faced challenges in maintaining content accuracy and speaker identity. These challenges underline the inherent complexity of voice conversion tasks, particularly when transferring speech style and content across speakers. 

Despite these advancements, controlling accents in voice conversion remains a challenge. The Singing Voice Conversion Challenge 2023 [8] made significant strides using a HuBERT-based feature extractor for linguistic content extraction and a DSPGAN vocoder for high-quality waveform generation. However, accent control was not a focus, which limits its broader applicability in tasks where accent modification is critical. 

Accent conversion, as demonstrated by various neural style transfer techniques [9], can bridge this gap, transforming non-native to native-like speech with higher accuracy. Our study uses the Speech Accent Archive, a parallel dataset, to control accents in our voice conversion model. While parallel data has its limitations, the availability of this dataset allows us to leverage the benefits of paired data. Previous research [10] has shown that parallel voice conversion techniques can still be effective with limited training data, using stochastic variational deep kernel learning (SVDKL) to balance model complexity and data fitting. Although we employ a different model (HuBERT-HiFi-GAN), this work informs our approach by demonstrating the feasibility of parallel techniques even with minimal data.

\section{Methodology}
\label{sec:methodology}

\subsection{Problem Statement}
The central issue addressed in this research is the challenge of achieving high-quality voice conversion, particularly for singing and accented speech, while overcoming the limitations of existing systems. Despite progress, achieving high-quality voice conversion while retaining important attributes such as emotional depth, accent control, and non-verbal sound elements remains difficult. Parallel voice conversion methods rely on paired datasets, which limits their applicability, while non-parallel methods have difficulty maintaining the full range of vocal and acoustic features. Moreover, voice conversion systems often fail to address accent transformation, especially for non-native speakers, which is critical in voice dubbing and speech synthesis applications. 

To overcome these challenges, we propose a model that leverages a HuBERT-based encoder for parallel data and non-parallel data training and a HiFi-GAN vocoder for high-quality audio generation. This approach ensures natural voice conversion, retaining essential vocal features while allowing for accent and pitch control. 

\subsection{Proposed Methodology}
The proposed methodology integrates a two-stage training process designed to perform high-quality voice conversion for both singing voice and accented speech. The core architecture of the model combines a HuBERT encoder-decoder framework with a HiFi-GAN vocoder. The methodology leverages both non-parallel data for singing voice conversion and parallel data for accent transformation while preserving the core acoustic properties such as timbre, pitch, and prosody. 

\begin{figure} % 't!' forces the figure to the top of the column
    \centering
\includegraphics[width=\textwidth,height=\textheight,keepaspectratio]{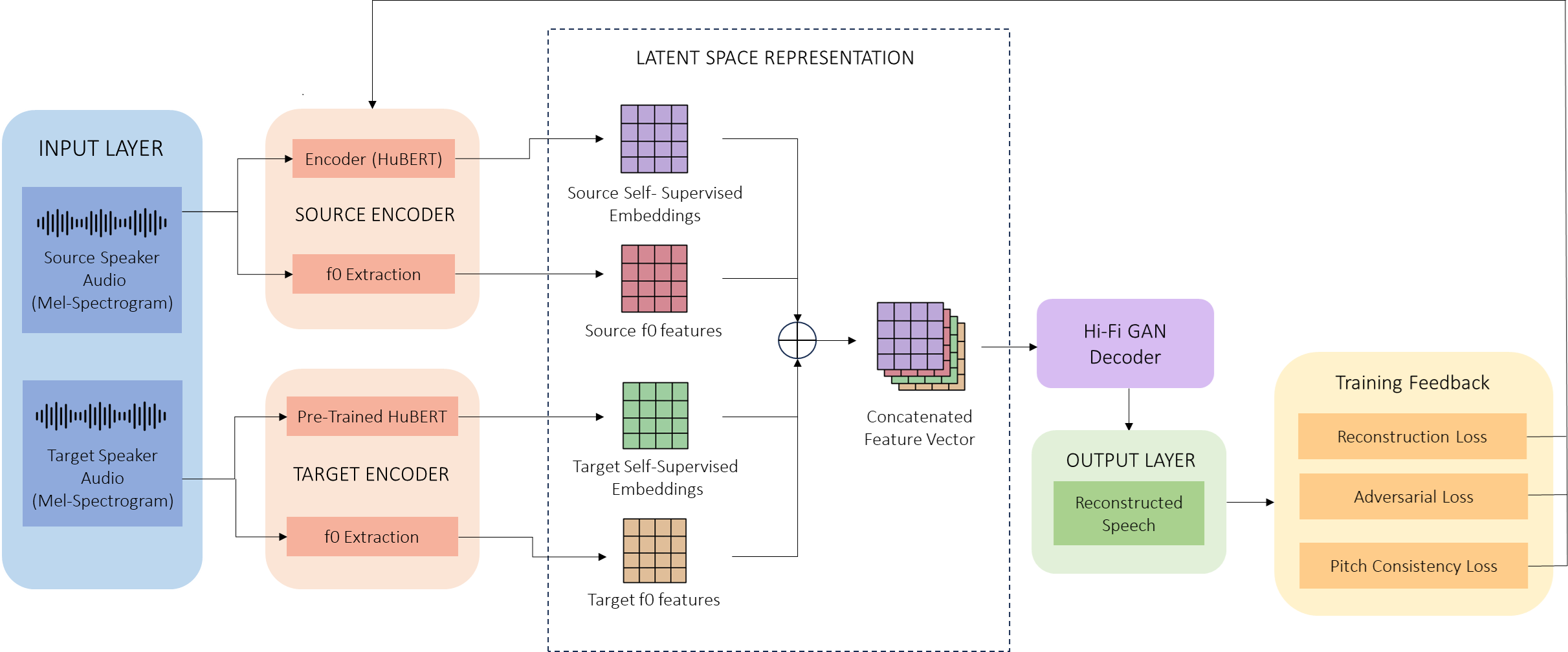}
    \caption{
        Architecture of the proposed voice conversion model
    }
    \label{fig3}
\end{figure}

The first stage of the training utilizes a non-parallel dataset that consists of recordings from seven singers (four female and three male), with approximately three hours of singing audio and 30 minutes to one hour of speaking audio per singer. These audio files are pre-processed by slicing them into 5-second clips to maintain consistency in input size, resampled to 44.1 kHz, and converted to mono. The processed audio is then transformed into mel-spectrograms, which serve as the input for the model. The mel-spectrogram provides a time-frequency representation of the audio, which is essential for capturing musical nuances and vocal tonalities. 

Following this, the second stage fine-tunes the model using a parallel dataset from the Speech Accent Archive [11]. This dataset includes English voice samples categorized by regional accents such as USA (en-US), UK (en-GB), Australia (en-AU), and Jamaica (en-JM). The parallel data allows the model to map specific phonetic and prosodic patterns across accents, facilitating the conversion of speech accents while retaining content. 

The HuBERT-based encoder is the first key component in the architecture. It operates on mel-spectrogram inputs and extracts meaningful features through a combination of 1D convolutional layers and multi-head self-attention layers. The encoder consists of 12 transformer layers, each with a hidden size of 768 and 12 attention heads. These multi-head self-attention layers capture long-term dependencies across time steps in the audio sequence, while the 1D convolutional layers capture local time-frequency patterns and ensure smooth feature extraction across the time axis. 

In this model, the source audio is first passed through the HuBERT encoder, which we train to extract self-supervised embeddings. These embeddings represent the content of the source audio, capturing the phonetic information while discarding speaker identity. Alongside this, a separate process extracts the source f0 (fundamental frequency) features, which are critical for pitch control. This is particularly useful for singing voice conversion, where precise pitch manipulation is required to preserve musicality. 

The target audio undergoes a parallel process. A pre-trained HuBERT encoder extracts self-supervised embeddings from the target speaker's audio, providing additional voice characteristics. If this step is omitted, the system relies solely on the f0 features from the target audio. Target f0 features are also extracted, enabling the system to adopt the prosody (pitch contour) of the target speaker, ensuring that the converted voice aligns with the pitch dynamics of the target. 

Once the source and target embeddings, along with their respective f0 features, are extracted, they are concatenated into a single feature vector. This concatenation step is crucial, as it combines the source's content information with the pitch and prosodic features of both the source and target. Specifically, the following features are concatenated: 
\begin{enumerate}
  \item Source self-supervised embeddings
  \item Source f0 features
  \item Target self-supervised embeddings 
  \item Target f0 features
\end{enumerate}

his concatenated feature vector encapsulates a rich combination of the source content, the pitch dynamics of both speakers, and, when included, the voice characteristics of the target. 

 \smallskip

After feature concatenation, the architecture transitions to the decoding phase, where a HiFi-GAN vocoder synthesizes the final audio output. The HiFi-GAN vocoder, a Generative Adversarial Network (GAN), generates high-quality audio from the concatenated feature vector. It consists of three up-sampling layers with a 4x4 kernel to progressively upscale the feature vector into the waveform domain. HiFi-GAN's use of multiple discriminators operating at different time scales ensures that both short-term and long-term dependencies are captured, making the synthesized output natural and coherent. The vocoder operates at over 100x real-time, allowing for efficient generation of high-fidelity speech and singing voice conversion output. 

The model's performance is optimized through several loss functions that provide training feedback. The first is the Reconstruction Loss, which measures the difference between the predicted and target mel-spectrograms using the Mean Squared Error (MSE). This loss ensures that the spectral features of the generated audio match those of the target audio. The formula for MSE loss is: 

\begin{equation}
L_{\text{reconstruction}} = \frac{1}{N} \sum_{i=1}^{N} \left( \hat{y}_i - y_i \right)^2
\label{eq:mse_loss}
\end{equation}

where  \( \hat{y}_i \) represents the predicted mel-spectrogram, 
\( y_i \) represents the target mel-spectrogram, and \( N \)
 is the number of frames. 

\medskip 

The second key component is Adversarial Loss, derived from the GAN framework. It ensures that the audio generated by the HiFi-GAN vocoder is perceptually convincing. The discriminator evaluates whether the generated audio is real or fake, and the generator is trained to fool the discriminator. The adversarial loss is computed as follows: 

\begin{equation}
L_{\text{adv}} = \mathbb{E} \left[ \left( D(\hat{x}) - 1 \right)^2 \right] + \mathbb{E} \left[ \left( D(x) - 0 \right)^2 \right]
\label{eq:adv_loss}
\end{equation}

where \( D(\hat{x}) \) is the discriminator’s prediction for the generated audio and \( D(x) \) is the discriminator’s prediction for the real audio.

\medskip 

Finally, the Pitch Consistency Loss ensures that the f0 features of the generated and target voices align. This is critical for singing voice conversion, where maintaining pitch accuracy is essential. The model minimizes the L1 norm between the predicted and target f0 values. The formula for pitch consistency loss is:

\begin{equation}
L_{\text{pitch}} = \frac{1}{N} \sum_{i=1}^{N} \left| f0_{\text{pred}, i} - f0_{\text{target}, i} \right|
\label{eq:pitch_loss}
\end{equation}

where \( f0_{\text{pred}, i} \) is the predicted pitch and \( f0_{\text{target}, i} \) is the target pitch at time step \( i \).

\medskip 

After training, the model effectively converts speech or singing into any of the target voices, maintaining high fidelity in pitch, timbre, and content reproduction.

\section{Result Analysis}
\label{sec:resultanalysis}

\subsection{Voice and Accent Classification}

\begin{table}[h]
\caption{Voice and Accent Classification Test}
\centering
\begin{tabular}{lcccc}
\toprule
\textbf{Voice \& Accent Approaches} & \textbf{HuBERT} & \textbf{HuBERT-Inter} & \textbf{ContentVEC} & \textbf{Proposed Model} \\
\midrule
\textbf{Voice Identification}  & 73.7 & 73.4 & 37.7 & 90.6 \\
\textbf{Accent Classification} & 81.6 & 72.7 & 62.1 & 95.6 \\ 
\bottomrule
\end{tabular}
\label{tab1}
\end{table}

The table I presents a comparative analysis of different voice and accent conversion approaches, namely HuBERT, HuBERT-Inter, ContentVEC, and the Proposed Model, based on their performance in voice identification and accent classification tasks. For voice identification, the accuracy scores are as follows: HuBERT achieved 73.7\%, HuBERT-Inter scored 73.4\%, ContentVEC demonstrated 37.7\%, while the Proposed Model attained a significantly higher accuracy of 90.6\%. In the accent classification task, the models' accuracy scores reflect their ability to differentiate accents, with HuBERT at 81.6\%, HuBERT-Inter at 72.7\%, ContentVEC at 62.1\%, and the Proposed Model achieving an impressive 95.6\%. These results indicate that the Proposed Model surpasses the other approaches in both tasks, underscoring its superior capability in capturing the nuances of both voice and accent features. 

\begin{figure} % 't!' forces the figure to the top of the column
    \centering
    \includegraphics[width=0.7\linewidth]{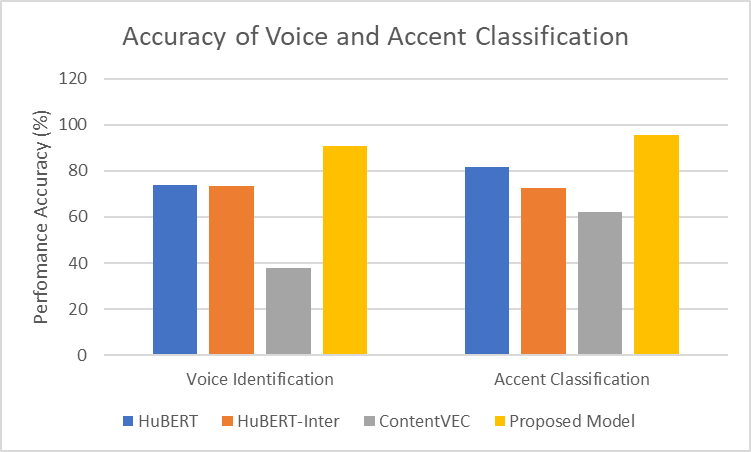}  % Adjust image width as needed
    \caption{
       Representation of Voice and Accent Conversion Test
    }
    \label{fig4}
\end{figure}

In Fig. 4, a representation of the performance comparison among different voice and accent approaches is depicted in a bar chart. The chart illustrates the accuracy scores for voice identification and accent classification tasks. In the bar chart, each approach (HUBERT, HUBERT-Inter, ContentVEC, and Proposed Model) is represented by a pair of bars, with one set for voice identification and another set for accent classification. The bar heights correspond to the respective accuracy scores, allowing for a quick visual comparison of the performance of each approach in the specified tasks. Notably, the Proposed Model demonstrates superior performance in both voice identification and accent classification, outperforming the other approaches.

\subsection{Voice Conversion Test Performance}

\begin{table}[h]
\caption{Described Voice Conversion Test Performance}
\centering
\begin{tabular}{lcccc}
\toprule
\textbf{Voice Conversion Method} & \textbf{C2C} & \textbf{O2C} & \textbf{C2O} & \textbf{O2O} \\
\midrule
\textbf{HuBERT}          & 0.9316 & 0.9277 & 0.9150 & 0.9257 \\
\textbf{HuBERT-Inter}    & 0.9286 & 0.9243 & 0.9050 & 0.9215 \\
\textbf{ContentVEC}      & 0.9029 & 0.8982 & 0.8848 & 0.9036 \\
\textbf{Proposed Model}  & 0.9512 & 0.9428 & 0.9392 & 0.9432 \\
\bottomrule
\end{tabular}
\label{tab2}
\end{table}

\begin{figure} % 't!' forces the figure to the top of the column
    \centering
    \includegraphics[width=0.7\linewidth]{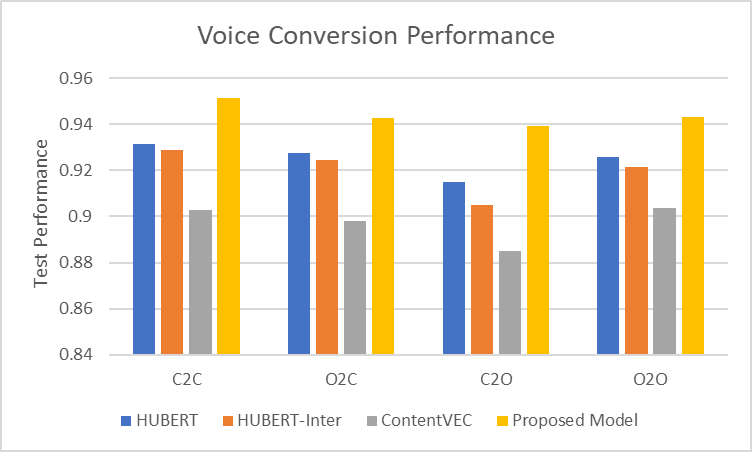}  % Adjust image width as needed
    \caption{
        Different Models’ Performance during Voice Conversion Test 
    }
    \label{fig5}
\end{figure}

The table II presents a comprehensive overview of the performance metrics for different Voice Conversion Methods across various conversion scenarios. The methods include HUBERT, HUBERT-Inter, ContentVEC, and Proposed Model, while the conversion scenarios consist of C2C (Conversion to Same Speaker), O2C (Original to Converted Speaker), C2O (Converted to Original Speaker), and O2O (Original to Original Speaker). 

In the bar chart (Fig. 5), each Voice Conversion Method (HUBERT, HUBERT-Inter, ContentVEC, and Proposed Model) is represented by a group of bars, with each bar corresponding to a specific conversion scenario (C2C, O2C, C2O, O2O). The height of each bar indicates the respective performance score for the method in the given conversion scenario. The numerical values in the table represent the performance scores, from an evaluation metric, for each method in the respective conversion scenarios. 

For instance, in the C2C scenario, HUBERT achieves a performance score of 0.9316, while HUBERT-Inter scores 0.9286, ContentVEC achieves 0.9029, and Proposed Model attains the highest performance with a score of 0.9512. Notably, the Proposed Model consistently exhibits higher performance scores across all scenarios, showcasing its effectiveness in voice conversion tasks. 

\section{Conclusion}

The proposed model demonstrates a significant advancement in voice conversion techniques by effectively integrating components such as HuBERT as the speech encoder and HiFi-GAN as the vocoder. This architecture effectively facilitates not only the transformation of source audio into various target voices but also handles complex accent conversions. By leveraging fundamental frequency (f0) features and self-supervised embeddings, the model ensures that both voice characteristics—like pitch and timbre—and accent features are preserved and accurately rendered. This innovative architecture facilitates the conversion of source audio into diverse target voices, while preserving essential voice characteristics like pitch and timbre. The system’s ability to handle complex voice transformations positions it as a versatile and powerful tool for various applications, including speech synthesis and voice dubbing. 

Future work could focus on refining this approach by exploring additional datasets and incorporating more sophisticated techniques to further enhance the versatility and performance of the voice conversion system. Such advancements would expand its applicability and effectiveness across a broader range of voice transformation tasks.

\end{document}